%% file: main.tex
\documentclass[%
preprint, 3p,
]{elsarticle}

\usepackage{siunitx}
\newcommand{\pd}[2]{\frac{\partial #1}{\partial #2}}

\usepackage{graphicx}
\usepackage{dcolumn}
\usepackage{bm}
\usepackage{xcolor}
\usepackage{amsmath}

\begin{document}
\begin{frontmatter}

\title{Theory of shock electrodialysis I: \\
Water dissociation and 
electrosmotic vortices }

\author[first]{Huanhuan Tian}
\author[first]{Mohammad A. Alkhadra}%
\author[first, second]{Martin Z. Bazant}
\ead{bazant@mit.edu}
\address[second]{Department of Mathematics, Massachusetts Institute of Technology. Massachusetts 02139, USA}

\address[first]{%
 Department of Chemical Engineering, Massachusetts Institute of Technology, Massachusetts 02139, USA
}%




\date{\today}

\begin{abstract}
Shock electrodialysis (shock ED), an emerging electrokinetic process for water purification, leverages the new physics of deionization shock waves in porous media.  In previous work, a simple leaky membrane model with surface conduction can explain the propagation of deionization shocks in a shock ED system, but it cannot quantitatively predict the deionization and conductance (which determines the energy consumption), and it cannot explain the selective removal of ions in experiments. This two-part series of work establishes a more comprehensive model for shock ED, which applies to multicomponent electrolytes and any electrical double layer thickness, captures the phenomena of electroosmosis, diffusioosmosis, and water dissociation, and incorporates more realistic boundary conditions. In this paper, we will present the model details and show that hydronium transport and electroosmotic vortices (at the inlet and outlet) play important roles in determining the deionization and conductance in shock ED. We also find that the results are quantitatively consistent with experimental data in the literature. Finally, the model is used to investigate design strategies for scale up and optimization. 

\end{abstract}

\end{frontmatter}



\input{1_intro}
\input{2_1_3D_model}

\input{2_2_DA_model}

\input{4_numerical}

\input{5_discussion}
\section*{Acknowledgments}

The research was supported by a Graduate Student Fellowship awarded by the MIT Abdul Latif Jameel Water and Food Systems Lab and funded by Xylem, Inc. The authors would like to thank Mohammad Mirzadeh and Pedro de Souza for insightful discussions.

\input{appendix}

\bibliographystyle{elsarticle-num} 
\bibliography{references, references_manual}

\end{document}

%% file: 1_intro.tex
\section{Introduction}

The ability to efficiently remove ions and ionic compounds from a dilute feed is central to a sustainable future with clean water. With growing industrial development, for example, toxic heavy metals, radioactive ions, and inorganic compounds are increasingly discharged into the environment and into our sources of drinking water. These contaminants are hazardous even when present in trace quantities, and it remains a challenge to remove them affordably and reliably \cite{Shannon2008ScienceDecades, Fu2011RemovalReview}. This challenge creates an urgent need for the development of systems by which charged species are selectively separated from dilute feeds. Traditional membrane processes for water treatment such as reverse osmosis (RO) and electrodialysis (EO)---which have proven successful for desalination of concentrated solutions like seawater---are not selective and have limited utility when the objective is to remove target ions in the presence of a competing electrolyte like sodium \cite{Kim2012CompetitiveProperties, VanDerBruggen2004SeparationNanofiltration}. 

\begin{figure} [h]
    \centering
    \includegraphics[width=0.45 \textwidth]{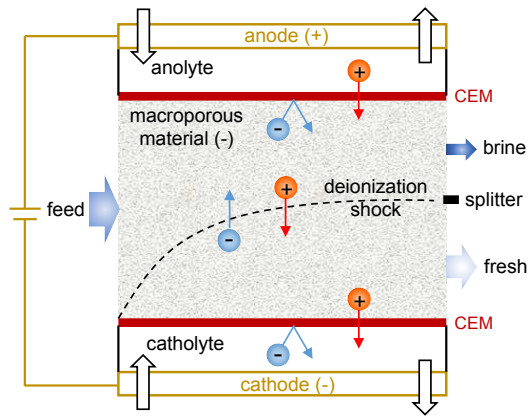}
    \caption{Schematic of the shock ED system comprising a negatively charged microporous frit between two cation exchange membranes (CEMs).}
    \label{fig:prototype}
\end{figure}

Over the past decade, Bazant and coworkers have developed, patented, and characterized an emerging electrokinetic process called ``shock electrodialysis (shock ED)'' for water purification \cite{Mani2011DeionizationMicrostructures, Deng2013OverlimitingMedia, Deng2015WaterDisinfection, Schlumpberger2015ScalableElectrodialysis, Bazant2014, Bazant2015DesalinationSystem}. As shown in Fig.\ref{fig:prototype}, the main component of the shock ED prototype is a weakly charged macroporous material sandwiched between two ion exchange membranes. In the current prototype, the macroporous material and both ion exchange membranes have negatively charged surfaces. When current is applied to the system, the membranes block the transport of coions (i.e., anions) and cause ion concentration polarization, which leads to extreme concentration gradients in the macroporous material. As the concentration is reduced to nearly zero on the depleted side, the current reaches the so-called diffusion limited current. However, overlimiting current is widely observed in electrochemical systems, and possible mechanisms include water splitting, electro-convective instability, surface conduction, and electroosmotic flow \cite{Dydek2011OverlimitingMicrochannel, Nikonenko2014DesalinationPerspectives}. In the shock ED prototype, because the macroporous material has surface charge and pore size of around $\SI{1}{\mu m}$, surface conduction and electroosmotic flow should dominate overlimiting current, which enables a deionization shock wave that propagates from one membrane to the other \cite{Dydek2013NonlinearModel, Mani2011DeionizationMicrostructures, Nam2015ExperimentalMicrochannels, Yaroshchuk2012Over-limitingAnalysis}. A contaminated stream fed to our shock ED device can therefore be split into fresh and waste products at the outlet. This process is continuous, scalable, and potentially free of membranes (for example by replacing the ion exchange membranes with nanoporous ceramics), making it promising for water treatment \cite{Wenten2020NovelProcesses}.

Schlumpberger et al. built and tested the first shock ED device and achieved over 99\% removal of Na$^{+}$ and K$^{+}$ from binary electrolytes with water recovery as high as 80\% \cite{Schlumpberger2015ScalableElectrodialysis}. Very recently, the Bazant group also proved the selective ion removal from electrolyte mixtures by shock ED \cite{Conforti2019ContinuousElectrodialysisb, Alkhadra2019ContinuousElectrodialysis, Alkhadra2020Small-scaleElectrodialysis}, which will be discussed in more detail in the second part of the series paper. Despite the significance of these recent experimental results, there remain open questions in the underlying theory that must be addressed by mathematical modeling. Dydek et al. were the first to describe the bi-ion shock ED system with plug crossflow using a simple homogenized model \cite{Dydek2013NonlinearModel}, which assumes thin electrical double layer (EDL) and may be invalid in the ion-depleted zone. Schlumpberger et al. then introduced linear electroosmotic flow to the system \cite{Schlumpberger2020shockEDcross} and compared the model to experimental data \cite{Schlumpberger2015ScalableElectrodialysis}. This work captured to some extent the trends of desalination and water recovery observed experimentally, but it significantly overestimated desalination and underestimated the overlimiting conductance (which determines energy consumption). Possible reasons for this discrepancy include oversimplification of boundary conditions, the neglect of water ion transport, and the assumption of thin EDLs. The microscopic Poisson-Nernst-Plank-Stokes model, which is applicable for any EDL thickness and considers surface conduction, electroosmosis, and diffusioosmosis, has been used for ion depletion in a single microchannel \cite{Nielsen2014ConcentrationMicrochannel, Mani2009OnAnalysis} and pore-networks \cite{Alizadeh2017MultiscaleDerivation, Alizadeh2017MultiscaleApplications, Alizadeh2019ImpactMedia} at overlimiting current. However, it has never been applied to the shock ED system with cross flow.
Finally, there has also been no theoretical work demonstrating the selective removal of multivalent ions by shock ED, for which models more robust than the homogenized model may be required.

In this paper, we present a comprehensive model for  multiple-ion, planar shock ED systems, as shown in Fig.\ref{fig:model}. Salt ions as well as hydronium and hydroxide ions are all included in the system. Instead of the macroporous material, this system involves either a single charged channel or a stack of these channels. The system can therefore be described by a three-dimensional pore-scale Poisson-Nernst-Planck-Stokes model, which is applicable for any EDL thickness and can be integrated to produce the depth-averaged model under the assumption of thin channels. The depth-averaging strategy used in this paper is similar to the area-averaging in Ref. \cite{Mani2009OnAnalysis,  Peters2016AnalysisNanopores}. The shock ED system in this work also considers the electrode streams and ion exchange membranes because the assumption of ideal membranes does not provide enough boundary conditions when dealing with multiple ions. Lastly, this model system also includes the unsupported electrolyte (i.e., a matrix with no charge) at the inlet and outlet, which can capture the streaming potential as well as regulate the flow and transport in the charged channel.


%% file: 2_1_3D_model.tex
\section{Depth-averaged model for planar shock ED system}
\begin{figure}
    \centering 
    \includegraphics[width = 0.5 \textwidth]{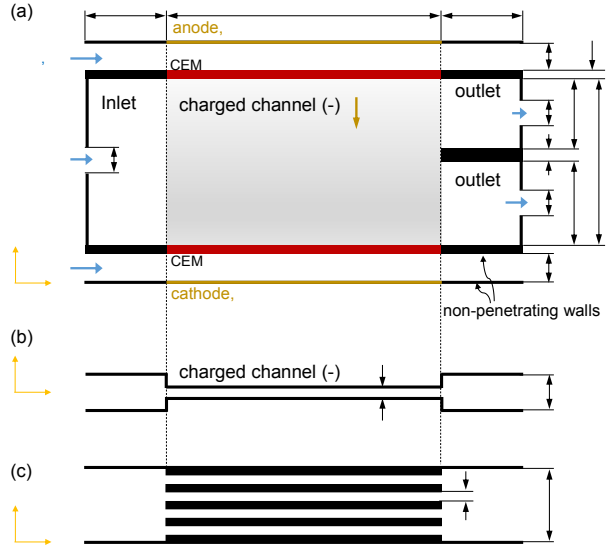}
    \caption{Two systems (a single channel and a stack) used in this paper to model shock ED. (a) Projection of both systems in the $xy$-plane, (b) projection of the single-channel in the $xz$-plane, and (c) projection of the stack in $xz$-plane.}
    \label{fig:model}
\end{figure}

{\label{sec:DA}}
In this section, we will provide a depth-averaged model for multiple ion transport in the single-channel system shown in Fig.\ref{fig:model}(a)(b). The system mimics the shock ED prototype in Fig.\ref{fig:model}, but it uses a planar, charged channel instead of the macroporous material. Two uncharged microchannels (one as the inlet and the other as the outlet) are connected to the charged channel, which allows us to use pore-scale equations to describe the transport in the feed stream, as presented in Sec.\ref{sec:general_eq}. We will then simplify the model under the assumption of thin channels in Sec.\ref{sec:thin_channel}, derive the depth-averaged equations in Sec.\ref{sec:DA_eq}, and give the boundary conditions of the feed channel in Sec.\ref{sec:B.C.}. Finally we will nondimensionalize the model in Sec.\ref{sec:non_dimen} and extend the model for a stack of charged channels (Fig.\ref{fig:model}(c)) in Sec.\ref{sec:stack}.


\subsection{General equations}{\label{sec:general_eq}}

Shock ED is a nonlinear process, where the ion concentration for each species $c_1$, $c_2$, $\cdots c_N$,  the flow velocity $\mathbf{u}=\{u, v, w\}$, the electric potential $\psi$, and the hydraulic pressure $p$ are coupled. Here $N=N_s+2$ is the total number of species, and we denote salt ions by subscripts $k=1,2,\cdots, N_s$, $\mathrm{H^+}$ by $k = N_s + 1$, and $\mathrm{OH^-}$ by $k = N_s + 2$. There are $N+5$ variables in total for a three-dimensional system, and we need the same number of equations to solve for them.

To begin with, ions are transported by convection, diffusion, and electromigration. For species $k$ ($k = 1, 2, \cdots, N$ unless specified otherwise) in a dilute solution, ion flux $\mathbf{J}_k$ can be described by the Nernst-Plank equation
\begin{equation}
    \mathbf{J}_k =  \mathbf{u} c_k - D_k \left(\nabla c_k + \frac{z_kc_k}{V_T} \nabla \psi \right),
    \label{eq:NPflux}
\end{equation}
where $D_k$, $z_k$ are respectively the diffusivity and valence of the species $k$, $V_T = k_b T/e$ is the thermal voltage where $k_b$ is the Boltzmann constant, $T$ is the temperature, and $e$ is the electron charge. We can add the ion fluxes together to get the electric current density:
\begin{equation}
\mathbf{i} = \sum_{k=1}^N z_k F \mathbf{J}_k,
\end{equation}
where $F$ is the Faraday constant. The conservation of species $k$ leads to
\begin{equation}
    \frac{\partial c_k}{\partial t} + \nabla \cdot \mathbf{J}_k = R_k, 
\end{equation}
where $R_k$ is the volumetric reaction rates of species $k$. In this work, we assume no reaction for salt ions, and only water dissociation reaction for water ions. So
\begin{equation}
    \frac{\partial c_k}{\partial t} + \nabla \cdot \mathbf{J}_k = 0 \ \ \{k = 1,2,\cdots, N_s\},
    \label{eq:NP}
\end{equation}
\begin{equation}
    \frac{\partial c_w}{\partial t} + \nabla \cdot \mathbf{J}_w = 0 ,
    \label{eq:NP_w}
\end{equation}
where $c_w = c_{N_s+1} - c_{N_s + 2}$ and $\mathbf{J}_w = \mathbf{J}_{N_s+1} - \mathbf{J}_{N_s+2}$ are respectively the difference of concentration and flux of the two water ions (H$^+$ and OH$^-$). Assuming fast dissociation of water,
\begin{equation}
    c_{N_s+1}c_{N_s+2} = K_w, 
    \label{eq:water_diss}
\end{equation}
where $K_w$ is the water dissociation constant. 

On the other hand, the flow is not only driven by a pressure gradient, but also by an electric force. A typical shock ED system involves an incompressible fluid and operates at low Reynolds number, hence the Stokes equations can be applied for momentum conservation: 
\begin{equation}
    \mu \nabla^2 \mathbf{u} - \nabla p - \rho_e \nabla \psi = 0,
\end{equation}
where $\rho_e$ is the local charge density. In addition, the continuity equation should be satisfied:
\begin{equation}
    \nabla \cdot \mathbf{u} = 0.
    \label{eq:continuity}
\end{equation}

Finally, the electric potential should satisfy the Poisson equation:
\begin{equation}
    \epsilon \nabla^2 \psi  = -\rho_e = -\sum_{k=1}^{N} z_k F c_k,
    \label{eq:Poisson}
\end{equation}
where $\epsilon$ is the permittivity of the solution and is assumed to be a constant. 

To summarize, Eq.(\ref{eq:NP})-(\ref{eq:Poisson}) provide $N+5$ equations so the system should be solvable with proper boundary conditions.

\subsection{Thin channel assumption}{\label{sec:thin_channel}}
In this part, we will simplify the full model in Sec.\ref{sec:general_eq} under the assumption of thin channels. The assumption includes two aspects. First, the $z$-dimension (less than $\SI{10}{\mu m}$) is much smaller than the $x$- and $y$-dimensions ($\sim \SI{}{mm}$), and thus we can assume $|\pd{}{z}|\gg |\pd{}{x}|, |\pd{}{y}|$ and parallel flow ($w=0$), which means that we can neglect the electroosmotic instability and electroosmotic vortices in $xz$ or $yz$ planes \cite{Dydek2011OverlimitingMicrochannel, Andersen2017ConfinementInstability}. Second, the P\'{e}clet number Pe$_k=|\mathbf{u}| h/D_k$, which represents the convection in the $x$- and $y$-directions over the diffusion in $z$-direction, is so small ($\ll 1$) that we can neglect Taylor-Aris dispersion \cite{Yaroshchuk2011CoupledApplicability} and assume local equilibrium ($J_{z,k} = 0$) to a virtual reservoir (where $\rho_e=0$) at each $(x, y)$ position \cite{Fair1971ReverseMembranes, Peters2016AnalysisNanopores}. We define $c_k^v(x,y)$, $\psi^v(x,y)$, $p^v(x,y)$ as the concentration, electric potential, and pressure in the virtual reservoirs. In the following, we will denote $\mathbf{u} = (u, v)$ and $\mathbf{J_k} =(J_{x,k}, J_{y,k})$ since $w=0$ and $J_{z,k}=0$, and we will denote $\tilde{\psi} = \psi/V_T$ for simplicity.

Firstly, we can specify the boundary conditions at the symmetric plane $z=0$ and the channel wall $z=h$: 
\begin{subequations}
\begin{equation}
    \left. \pd{\mathbf{u}}{z} \right|_{z=0}=\mathbf{0}, \ \left. \mathbf{u}\right|_{z=h} = \mathbf{0},
\end{equation}
\begin{equation}
    \left.\pd{\psi}{z}\right|_{z=0}=0,\ \left.  \pd{\psi}{z}\right|_{z=h}= E_s = \frac{\sigma}{\epsilon},
    \label{eq:bc_z_phi}
\end{equation}
\label{eq:bc_z}
\end{subequations}
where $h$ is the inverse of internal area density (channel surface area $a_p$ per channel volume $V_p$), i.e., half channel depth for planar channels, $\sigma$ is the surface charge density, and $E_s$ is the electrical field on the walls. In reality, $\sigma$ comes from surface group dissociation, which is dependent on the local ionic composition \cite{Behrens2001TheSurfaces}. The charged walls attract counter-ions and repel co-ions, and form  the Stern layer where ions are immobile and then the diffuse layer where ions are mobile \cite{Schoch2008TransportNanofluidics}. These space charge layers and the charged wall form the so called EDL. In this paper, we will fix $\sigma$ for simplicity, and neglect the Stern layer. 

Next we will simplify the governing equations under the assumption of thin channels. Firstly, we can decompose the electric potential $\psi = \psi^v(x, y) + \varphi(x, y, z)$, \cite{Fair1971ReverseMembranes} where $\varphi$ is the equilibrium part of potential from the Boltzmann distribution:
\begin{equation}
    c_k = c_k^v \exp(-z_k \tilde{\varphi}),
    \label{eq:boltzmann}
\end{equation}
which is essentially the results of $J_{z, k}=0$ and $w=0$, and we assume $\tilde{\varphi}=0$ in the reservoir. Note that $\sum z_k c_k^v = 0$ since $\rho_e = 0$ in the virtual reservoir. We will denote the $\varphi$ at charged walls as $\zeta$. Plug Eq.(\ref{eq:boltzmann}) into the Nernst-Plank (NP) flux (Eq.(\ref{eq:NPflux})), we arrive at
\begin{equation}
    \mathbf{J}_k =  \mathbf{u} c_k - D_k c_k \nabla_{xy} \left(\ln c_k^v + z_k \tilde{\psi}^v   \right).
    \label{eq:NPflux_thin}
\end{equation}
Plug Eq.(\ref{eq:boltzmann}) into the water dissociation equation, we get
\begin{equation}
    c_{N_s+1}^v c_{N_s+2}^v = K_w.
    \label{eq:water_diss_v}
\end{equation}
Plug Eq.(\ref{eq:boltzmann}) into the Poisson equation and apply $|\pd{}{z}|\gg |\pd{}{x}|, |\pd{}{y}|$, we have the local Poisson-Boltzmann (PB) equation:
\begin{equation}
    \epsilon \pd{^2 \varphi}{z^2} = -\rho_e = -\sum_{k=1}^{N} z_k F c_k^v \exp(-z_k \tilde{\varphi}),
    \label{eq:PB}
\end{equation}
which can be solved with the known $c_k^v$ and the boundary condition Eq.(\ref{eq:bc_z_phi}). It can also be rearranged to
\begin{equation}
    \lambda_D^2\frac{d^2 \tilde{\varphi}}{d z^2}=-\sum_{k=1}^{N} \frac{z_k c_k^v}{\mathcal{I}} \exp(-z_k \tilde{\varphi}),
    \label{eq:PB2}
\end{equation}
where $\mathcal{I} = \frac{1}{2}\sum z_k^2 c_k^v$ is the ionic strength and $\lambda_D = \sqrt{\epsilon V_T/ F \mathcal{I}}$ is the Debye length and the characteristic length for EDL thickness \cite{Bazant2009TowardsSolutions, Schoch2008TransportNanofluidics}.

Then we can simplify the Stokes equations to
\begin{equation}
    \mu \pd{^2\mathbf{u}}{z^2} - \nabla_{xy}p - \rho_e \nabla_{xy}\psi= 0,
    \label{eq:stokes_xy}
\end{equation}
\begin{equation}
    -\pd{p}{z} - \rho_e \pd{\varphi}{z} = 0,
    \label{eq:stokes_z}
\end{equation}
By substituting Eq.(\ref{eq:PB}) in to Eq.(\ref{eq:stokes_z}) and integrating it, we get
\begin{equation}
     p  - p^v = V_T F \sum_{k=1}^N  \left[\exp(-z_k \tilde{\varphi}) - 1 \right] c_k^v.
\end{equation}
Then we substitute the above equation and Eq.(\ref{eq:PB}) into Eq.(\ref{eq:stokes_xy}) to arrive at
\begin{equation}
\begin{aligned}
    \mu \pd{^2 \mathbf{u}}{z^2} - \nabla_{xy} p^v &+ \epsilon \pd{^2 \varphi}{z^2} \nabla_{xy} \psi^v \\ 
    - V_T F &\sum_{k=1}^N \left[\exp(-z_k \tilde{\varphi}) - 1\right] c_k^v \nabla_{xy} \ln c_k^v  = 0,
\end{aligned}
\end{equation}
Since the above equation is linear in terms of $\mathbf{u}$, it can be integrated twice to explicitly express $\mathbf{u}(x,y,z)$ as
\begin{equation}
    \mathbf{u} = \mathbf{u}^P + \mathbf{u}^{EO} + \sum_{k=1}^N \mathbf{u}^{DO}.
\end{equation}
where 
\begin{subequations}
\begin{equation}
    \mathbf{u}^P = -\frac{h^2 - z^2}{2\mu}  \nabla_{xy} p^v
    \label{eq:uP}
\end{equation}
is the pressure-driven flow part of $\mathbf{u}$, 
\begin{equation}
    \mathbf{u}^{EO} = -\frac{\epsilon (\varphi-\zeta) }{\mu} \nabla_{xy}\psi^v
    \label{eq:uEO}
\end{equation}
is the electroosmotic part, and 
\begin{equation}
    \mathbf{u}^{DO}_k =  -\frac{V_T F h^2 }{\mu}  c_k^v \chi_k \nabla_{xy} \ln c_k^v 
    \label{eq:uDO}
\end{equation}
\end{subequations}
is the diffusioosmotic part, and $\chi_k(x,y,z)$ is a dimensionless integral defined in Appendix \ref{sec:appendix}.

Up to now, we have reduced the original $N+5$ three-dimensional unknowns $c_k$, $\psi$, $p$, $u$, $v$, $w$ to $N+2$ independent two-dimensional unknowns $c_k^v$, $\psi^v$, $p^v$ and one three-dimensional unknown $\varphi$, which largely simplifies the problem.

\subsection{Depth-averaged equations}{\label{sec:DA_eq}}
In this section, we will integrate the three-dimensional model under thin channel assumption (Sec.(\ref{sec:thin_channel})) in the $z$ direction to get the depth-averaged equations. In the following, we will drop the subscripts in $\nabla_{xy}$ to simplify the notations. 

First, we average the velocities (Eq.(\ref{eq:uP})-(\ref{eq:uDO})), and arrive at
\begin{equation}
    \overline{\mathbf{u}^{eff}_k} = \frac{\overline{\mathbf{u} c_k}}{ \overline{c_k}} = \beta^{P}_k \overline{\mathbf{u}^{P}} +  \beta_k^{EO} \overline{\mathbf{u}^{EO}} + \sum_l \beta^{DO}_{kl} \overline{\mathbf{u}^{DO}_l},
    \label{eq:u_eff_ave}
\end{equation}
where the coefficients $\beta_k^P$, $\beta_k^{EO}$, and $\beta_{kl}^{DO}$ account for the difference between the averaged convective flux and the product of the averaged velocity and concentration. Please check Appendix \ref{sec:appendix} for how to calculate them.The velocities can be obtained from the fields of pressure, electric potential, and concentration:
\begin{subequations}
\begin{equation}
    \overline{\mathbf{u}^{P}} = -k^P \nabla p^v,
    \label{eq:uP_ave}
\end{equation}
\begin{equation}
    \overline{\mathbf{u}^{EO}} = -k^{EO} \nabla \psi^v,
\end{equation}
\begin{equation}
    \overline{\mathbf{u}^{DO}_k} = -  k_k^{DO} \nabla \ln c_k^v,
    \label{eq:uDO_ave}
\end{equation}
\end{subequations}
where $k^P = \frac{h^2}{3\mu}$, $k^{EO} = -\frac{ \epsilon \zeta}{\mu} \alpha^{EO}$, and $k^{DO}_k =  \frac{V_T F h^2}{\mu} c_k^v \alpha_k^{DO}$ are the permeability for each flow mechanism, and the dimensionless coefficients $\alpha^{EO}$ and $\alpha^{DO}$ are defined in Appendix \ref{sec:appendix}. 

Next, we define $\delta_k =\frac{\overline{c_k}}{c_k^v}=\frac{1}{h}$ to relate $c_k^v$ and $\overline{c_k}$. The total averaged concentration $\overline{c_k}$ can thus be decomposed to the bulk concentration $c_k^v$, and the surface excess concentration $(\delta_k - 1)c_k^v$.
The average of the NP flux (Eq.(\ref{eq:NPflux_thin})) yields
\begin{equation}
    \overline{\mathbf{J}_{k}} = \left[ \overline{\mathbf{u}^{eff}_k}   - D_k  \nabla \left(\ln (\overline{c_k}/\delta_k) + z_k  \tilde{\psi}^v \right) \right] \overline{c_k}.
    \label{eq:NPflux_ave}
\end{equation}
This flux is a combination of the bulk convection $\overline{\mathbf{u}} c_k^v$, the bulk diffusion $D_k c_k^v \nabla \ln c_k^v$, the bulk conduction $D_k z_k c_k^v \nabla  \tilde{\psi}^v$, as well as the surface convection $\left(\overline{\mathbf{u}_k^{eff}} \delta_k - \overline{\mathbf{u}} \right)c_k^v$, the surface diffusion $D_k (\delta_k -1)c_k^v  \nabla \ln c_k^v$, and the surface conduction $D_k z_k (\delta_k - 1)c_k^v \nabla  \tilde{\psi}^v$. 

Then we can integrate the mass conservation equation (Eq.(\ref{eq:NP})(\ref{eq:NP_w})), water dissociation equation (Eq.(\ref{eq:water_diss_v})), continuity equation (Eq.(\ref{eq:continuity})), and PB equation (Eq.(\ref{eq:PB2})) in the $z$ direction, and arrive at
\begin{equation}
    \pd{(h \overline{c_k})}{t} + \nabla \cdot (h \overline{\mathbf{J}_k} ) = 0 \ \ \{k = 1,2,\cdots, N_s\},
    \label{eq:NP_ave}
\end{equation}
\begin{equation}
    \pd{(h \overline{c_w})}{t} + \nabla \cdot (h \overline{\mathbf{J}_w} ) = 0,
    \label{eq:Npw_ave}
\end{equation}
\begin{equation}
    \overline{c_{N_s+1}} \ \overline{c_{N_s+1}} = K_w \delta_{N_s+1} \delta_{N_s+2},
\end{equation}
\begin{equation}
    \nabla \cdot \left[h\left(\overline{\mathbf{u}^{P}} +\overline{\mathbf{u}^{EO}} +\sum_{k=1}^N \overline{\mathbf{u}^{DO}_k}  \right)\right] =0,
    \label{eq:continuity_ave}
\end{equation}
and
\begin{equation}
    \sum_{k=1}^N z_k  \overline{c_k} - c_s = 0.
    \label{eq:electroneutrality}
\end{equation}
where $c_s = -\sigma/hF$ as the volume-averaged negative surface charge density. This last equation is called the electroneutrality equation. Now we have $N+2$ equations (Eq.(\ref{eq:NP_ave})-(\ref{eq:electroneutrality})) in terms of $N+2$ independent variables $\overline{c_k}$, $p^v$, and $\psi^v$, which can be solved with proper boundary conditions and known coefficients $\delta_k$, $\alpha^{EO}$, $\alpha^{DO}_k$, $\beta^P_k$, $\beta^{EO}_k$, $\beta^{DO}_{kl}$. In fact, these coefficients are largely determined by how much of the channel is occupied by EDLs. Please see Appendix \ref{sec:appendix} for how we calculate the coefficients. In the shock ED experiments \cite{Schlumpberger2015ScalableElectrodialysis, Conforti2019ContinuousElectrodialysisb}, typically $\mathcal{I}=$ 0.01--$\SI{100}{mM}$, which corresponds to $\lambda_D=$ 1--$\SI{100}{nm}$, and the pore diameter (i.e., $4h = 4 V_p/a_p$ for cylindrical pores) is around $\SI{1}{\mu m}$ for the macroporous material and around $\SI{1}{nm}$ for the membrane. So the EDL can be thin, comparable, or thick compared with these two pore sizes. Next, we will discuss the limits of thin and thick EDLs in the depth-averaged model.

If the EDL is much thinner than the channel depth ($ \lambda_D \ll h$),  
\begin{subequations}
\begin{equation}
    \delta_k, \alpha^{EO}, \beta^P_k, \beta^{EO}_k, \beta^{DO}_{kl} \sim 1,
\end{equation}
\begin{equation}
    \alpha^{DO}_k \sim 0.
\end{equation}
\end{subequations}
Therefore, the pore-scale details are no longer important, and the depth-averaged model is reduced to the homogenized model \cite{Dydek2013NonlinearModel}, where $k^{EO} = -\epsilon \zeta/\mu$, $k^{DO}_k = 0$, and $\overline{\mathbf{u}^{eff}_k} =\overline{ \mathbf{u}}$. 

If the EDL is much thicker than the channel depth ($\lambda_D \gg h$), we can assume $\varphi \approx \zeta$, $|\tilde{\zeta}| \gg 1$, and $\frac{\sigma}{hF} \approx - \sum_{k=1}^N z_k c_k^v \exp(-z_k \tilde{\zeta})$ \cite{Tian2017ElectrokineticInterface}. Therefore,
\begin{subequations}
\begin{equation}
    \delta_k \sim \exp(-z_k\tilde{\zeta}),
\end{equation}
\begin{equation}
    \alpha^{DO}_k \sim \left\{ \begin{array}{rcl} &\frac{1}{3}\exp(-z_k\tilde{\zeta}) & \mathrm{if\ } z_k \tilde{\zeta}<0\\
    &-\frac{1}{3} & \mathrm{if\ } z_k \tilde{\zeta}>0
    \end{array} \right.
\end{equation}
\begin{equation}
    \alpha^{EO} \sim 0,
    \label{eq:alpha_EO_thick}
\end{equation}
\begin{equation}
    \beta^P_k, \beta^{EO}_k, \beta^{DO}_{kl} \sim 1.
\end{equation}
\end{subequations}
We can further investigate the scales of the velocities. $k^P \sim \mathcal{O}(h^2)$, and Eq.(\ref{eq:alpha_EO_thick}) indicates that the $k^{EO} \sim 0$. Assuming constant $\sigma$, we have $\max_k\exp(- z_k \tilde{\zeta}) \sim \mathcal{O}(h^{-1})$, and so $\max_k k^{DO}_k \sim \mathcal{O}(h)$. Therefore, the pressure-driven flow, electroosmosis, and diffusioosmosis all decrease as $h$ becomes smaller, while diffusion and electromigration are not much influenced by the pore size. To conclude, the flow can be neglected at the thick EDL limit and extremely small $h$. As a result, we can apply these scalings to membranes. 
In fact, under such extreme confinement, the electroneutrality may break down \cite{Levy2020BreakdownNanopores} and the ion transport may be correlated \cite{Faucher2019CriticalPerspective}. However, for typical ion concentrations ($\sim$ mM) and membrane charge (charge concentration $\sim$ M) in this work, the above effects should not be very important.

%% file: 2_2_DA_model.tex
\subsection{Boundary conditions}{\label{sec:B.C.}}

In this part we will give the boundary conditions for the feed channel (including the unsupported electrolyte zones) shown in Fig.\ref{fig:model}(a), which will be combined with the depth-averaged equations to complete the model. 

At the inlet, we assume
\begin{equation}
    \overline{u} = u^{in} ,\  \overline{c_k} = c_k^{in},\  \pd{\psi^v}{x} = 0.
\end{equation}
Note that since the inlet zone is not charged, $\sum z_k c_k^{in} = 0$ should be satisfied. 
At the outlet, we assume 
\begin{equation}
     p^v = p_a,\ \pd{\overline{c_k}}{x} =0,\ \pd{\psi^v}{x} = 0,
\end{equation}
where $p_a$ is the atmospheric pressure.



On all the non-penetrating walls shown in Fig.\ref{fig:model}(a) (solid black lines or blocks, e.g., the splitter), apply
\begin{equation}
    \mathbf{u}  \cdot \mathbf{n} = 0, \ \mathbf{J}_k \cdot \mathbf{n} = 0.
    \label{eq:non_penetration}
\end{equation}
where $\mathbf{n}$ is the normal vector of the walls pointing to the outside of the channel. On these walls the viscous boundary layer thickness should be comparable to the channel depth, which is much smaller than the $x$ and $y$ dimensions, so the non-slip boundary condition is not applied here.   

Finally, we need to give the upper and lower boundary conditions of the charged channel. When only two species exist (one anion, one cation), the previous work \cite{Dydek2013NonlinearModel, Schlumpberger2020shockEDcross} just assumes $J_{y, -} = 0$  (ideal CEMs) and $\psi^v = \pm V/2$ at the anode and cathode side membranes. However, this leads to almost perfect ion removal, which cannot be used to predict the magnitude of outlet concentration. Furthermore, these two boundary conditions are not enough for a multi-component system. A possible solution is to assign artificial transference numbers
to all the ions, but it turns out this could be too rigid of a boundary condition since transference numbers can depend on electric current. In this paper, we choose to directly simulate the transport in the membranes and electrode channels, so we will give boundary conditions in the electrode channel instead of on the upper and lower boundaries of the charged channel. We will use the depth-averaged equations in the limit of thick EDLs for transport in the membranes. However, since the membrane is porous, we need to modify the $D_k$ to $D_k^{eff}$. In this paper, we simply assume $D_k^{eff}/D_k = 0.1$ for any $k$. For the electrode channels, which are not charged, we use the homogenized equations and assume plug flow with $\overline{u} = u^E$. At the inlet, we assume $\pd{\psi^v}{x} = 0$, and $\overline{c_k} = c_k^{A}$ for the anode and $\overline{c_k} = c_k^{C}$ for the cathode stream. At the outlet, we assume $\pd{\overline{c_k}}{x} = 0$ and $\pd{\psi^v}{x} = 0$. On the non-penetrating walls, also assume Eq.(\ref{eq:non_penetration}). On the electrode surface, assume $\mathbf{u} \cdot \mathbf{n} = 0$, $\mathbf{J}_k \cdot \mathbf{n} = 0$ for $k = 1,2,\cdots, N_s$, and $\psi^v = \pm V/2$. Here we assume fast reaction on electrodes so the reaction rate is limited by ionic flux of water ions. Note that $V$ may differ from the true voltage between electrodes due to surface overpotentials.

\subsection{Non-dimensionalization}{\label{sec:non_dimen}}
In this part, we will non-dimensionalize the PB equation (Eq.(\ref{eq:PB2})), the depth-averaged equations (Eq.(\ref{eq:uP_ave})-(\ref{eq:uDO_ave}), (\ref{eq:u_eff_ave})-(\ref{eq:electroneutrality})) and the boundary conditions. The scales we choose to non-dimensionalize variables are summarized in Table \ref{tab:table1}. We will use the hat accent ($\hat{f}$) to represent the dimensionless averaged quantities, and use the tilde accent ($\tilde{f}$) to represent all the other dimensionless quantities. Except for the accents, the dimensionless equations and boundary conditions look the same as the dimensional form. Also note that by definition, the integrals $\alpha^{EO}$, $\alpha^{DO}_k$, $\beta^P_k$, $\beta^{EO}_k$, $\beta^{DO}_k$, $\delta_k$ themselves are dimensionless. Though we have used $h$ and $z$ to calculate the integrals, their dimensions cancels each other and the two variables can be substituted by $1$ and $\tilde{z}=z/h$ respectively to get the fully dimensionless form.
To compare with experiments, we also scale the total current by the advection of positive charges $\sum z_+ c_+ F u^{in} H_{io} h_{io}$, where $H_{io}$ and $h_{io}$ are respectively the width and depth of the inlet and outlet for the feed.
\begin{table}[b]
\caption{\label{tab:table1}%
A summary of the variables and parameters and their scales used in the depth-averaged model. In the context, we will use the hat accent ($\hat{f}$) to represent the dimensionless averaged quantities, and use the tilde accent ($\tilde{f}$) to represent all the other dimensionless quantities.
}
\centering
\begin{tabular}{c c}
\hline
variables or parameters & global dimensional scales\\ \hline
$D_k$ & $D_0 = \SI{1e-9}{m^2/s}$ \\
$h$ & $h_{io}$ \\
$x$,$y$ & $L$ \\
$t$ & $L^2/D_0$ \\
$\overline{c_k}$, $c_s$, $c_k^{in}$, $c_k^A$, $c_k^C$, $c^F_k$ & $c_0 = \sum_{salt} z_+ c_+^{in} $ \\
$\psi^v$, $\varphi$, $V$, $\zeta$ & $V_T = k_b T/e$ \\
$p^v$, $p^a$ & $p_0 = p_a/100$ \\
$\overline{\mathbf{u}}$, $u^{in}$, $u^E$ & $D_0/L$  \\ 
$\mathbf{J}_k$ & $D_0 c_0/L$\\
$k^P$ & $D_0/p_0$\\
$k^{EO}$ & $D_0/V_T$\\ 
$k^{DO}_l$ & $D_0$ \\
$K_w$ & $c_0^2$ \\
$I$ & $\sum z_+c_+F u^{in} H_{io} h_{io}$\\\hline
variables & local dimensional scales \\ \hline
$z$, $\lambda_D$ & $h$ \\ 
$E_s$ & $V_T/h$ \\\hline
\end{tabular}
\end{table}

\subsection{Extension for stack of channels}{\label{sec:stack}}

In this part, we will extend the depth-averaged model to stack of channels as shown in Fig.\ref{fig:model}. We assume $h_{io}\ll L, H$, which guarantees the validity of the thin channel assumption, and we further assume that the profiles of physical variables are identical in each charged channel. We only need to define a new parameter: porosity $\epsilon_p$, which is the sum of all the space of charged channels in the $z$ direction divided by $h_{io}$. Then we can just replace the $\tilde{h}$ by $\epsilon_p$ in the Nernst-Planck equation (Eq.({\ref{eq:NP_ave}}), (\ref{eq:Npw_ave})) and the continuity equation (Eq.(\ref{eq:continuity_ave})). This is similar to the model for porous media, but here we use planner channels so the tortuosity is 1. However, since we only care about the steady states, we can choose an artificial porosity to imitate the real porous media.

%% file: 4_numerical.tex
\section{Numerical simulation}{\label{sec:sim}}

\subsection{Algorithms and parameters}
In this section, we present the simulation results of the depth-averaged model. The model is solved by the following procedure. First, we solve the PB equation and tabulate $\alpha^{EO}$, $\alpha^{DO}_k$, $\beta^P_k$, $\beta^{EO}_k$, $\beta^{DO}_{kl}$, $\delta_k$, $\overline{c_k}$ and $\zeta$ in terms of $c_k^v$ for pre-selected values of $h$ and $\sigma$. We then discretize the depth-averaged equations using the finite volume method with nonuniform meshes because the membranes and electrode channels are much thinner in the $y$ direction relative to the feed channel. Next, we solve the transient depth-averaged equations until the steady-state solutions are obtained: (1) give the initial $\hat{c}_k^0$, $\tilde{\psi}^{v,0}$, and $\tilde{p}^{v, 0}$, (2) find the $c_k^v$ corresponding to $\overline{c_k}^0$,  (3) find the $\alpha^{EO}$, $\alpha^{DO}_k$, $\beta^P_k$, $\beta^{EO}_k$, $\beta^{DO}_{kl}$, $\delta_k$ and $\zeta$ based on $c_k^v$, (4) calculate the $\tilde{k}^P$, $\tilde{k}^{EO}$, and the diffuoosmotic flow $\hat{u}^{DO}_k$, (5) solve for the depth-averaged equations with fixed coefficients and $\hat{u}^{DO}_k$ and get the $\hat{c}_k$, $\tilde{\psi}^v$, $\tilde{p}^v$ for the new time step,  (6) end the process if the difference between $\hat{c}_k$, $\tilde{\psi}^v$, $\tilde{p}^v$ and $\hat{c}_k^0$, $\tilde{\psi}^{v,0}$, and $\tilde{p}^{v, 0}$ is smaller than a critical value, otherwise let $\hat{c}_k^0 = \hat{c}_k$, $\tilde{\psi}^{v,0} = \tilde{\psi}^v$, $\tilde{p}^{v, 0} = \tilde{p}^v$ and repeat steps (2)-(6). 

To better understand the mechanisms, we simulate not only the full depth-averaged model, but also the reduced depth-averaged model with flow driven only by a pressure gradient (no electroosmosis and diffusioosmosis), the full homogenized model, and the reduced homogenized model with flow driven only by a pressure gradient. The above four models are named as DAfull, DAp, Hfull, and Hp, respectively, in this paper (as shown in Table \ref{tab:model}). For given values of $\sigma$ and $h_p$, we solve the Grahame equation for a single EDL at $c_k^{in}$ to get a constant $\zeta$ for the Hfull model, and we solve the pore-scale profiles to get $\zeta$ as a function of $\overline{c_k}$ for the DAfull model. As a result, DAfull model has larger $|\zeta|$ in the ion-depleted zone. By comparing Hp and DAp, or Hfull and DAfull,  we can learn how the pore-scale details influence the results. Also, by comparing DAfull and DAp, we can determine the roles of electroosmosis and diffusioosmosis. In addition, we reproduce the model of Schlumpberger et al.\cite{Schlumpberger2020shockEDcross} and compare it with our models. 

\begin{table}
    \centering
    \caption{The notation for the models used in this paper.}
    \begin{tabular}{c|c}
        \hline
        DAfull & full depth-averaged model \\ \hline
        DAp & reduced depth-averaged model where \\
        & flow is only driven by pressure gradient\\ \hline
        Hfull & full homogenized model \\ \hline
        Hp & reduced homogenized model where \\
        & flow is only driven by pressure gradient\\ \hline
    \end{tabular}
    
    \label{tab:model}
\end{table}
\begin{table}[b]
\centering
\caption{\label{tab:constants}%
A summary of constants we use for the simulations.
}
\begin{tabular}{c c| c c |c}
\hline
\multicolumn{2}{c|}{$D_k$ ($10^{-9}\SI{}{m^2/s}$)} & \multicolumn{2}{c|}{Geometry (mm)} & Others \\ \hline
Na & 1.33 & $H_{io}$ & 0.6 & $\mu = \SI{8.9e-4}{Pa\cdot s}$\\
K & 1.96 & $H_E$ & 0.65 & $V_T = \SI{25.7}{mV}$ \\
Cl & 2.03 & $H_s$ & 1 & $\epsilon_w = \SI{6.947e-10}{F/m}$ \\
H & 9.31 & $H_M$ & 0.13 & $K_w = \SI{1e-8}{mM^2}$ \\
OH & 5.27 & $L_I$ & 3 & $c_s^{M} = \SI{2.63}{M}$ \\
 & & $L_O$ & 2  & $\epsilon_p = 0.33$ \\
 & & $L$ & 10 &  \\
  & &$h_{io}$ & $\SI{2.5e-3}{}$  & \\ \hline
\end{tabular}
\end{table}

Though the model is for planar channels, we can still choose parameters to imitate the experimental conditions. In this work, we aim to qualitatively and quantitatively compare the simulation results with the experiments by Schlumpberger et al. \cite{Schlumpberger2015ScalableElectrodialysis}. They used ultrafine borosilicate glass frits (Adams \& Chittenden Scientific Glass), which had a nominal pore size of $0.9$--$\SI{1.4}{\mu m}$ and a porosity of $48\%$. Therefore, the frit should have $h_p = V_p/a_p \approx 225$--$\SI{350}{nm}$ and a hindrance factor $\epsilon_p/\tau \approx \epsilon_p^{1.5} \approx 33\%$.  In this paper, we choose $h_p = \SI{250}{nm}$ and $\epsilon_p = 33\%$ for the stack of planar charged channels.
In addition, Schlumpberger et al. \cite{Schlumpberger2015ScalableElectrodialysis} used Nafion 115 (0.005 inches thick) as the membrane, which has 1 meq/g total acid capacity and 38\% water uptake (by weight). Since water density is about $\SI{1}{g/cm^3}$, the negative charge concentration in membrane $c_s^M \approx 1\times1/0.38 \approx \SI{2.63}{M}$. Nafion will also slightly expand in water, and the charge density can slightly vary with pH. In this paper, we simply fix $c_s = \SI{2.63}{M}$, $H_M =  \SI{130}{\mu m}$ for the membrane.
We set $h_{io}=\SI{2.5}{\mu m}$, which is apparently different from the depth of the real prototype ($W = \SI{2}{cm}$). However, we set $u^{in} = \frac{Q}{W H_{io}}$ and $u^{E} = \frac{Q_E}{W H_E}$, where $Q$ and $Q_E$ is the experimental flow rate into the feed and electrode channels, to make the flow rate per unit depth the same as experiments. Meanwhile, the dimensionless current $\tilde{I}$ is independent of the depth in $z$ direction. In addition, we set $H_{io} = \SI{0.6}{mm}$,$H_E = \SI{0.65}{mm}$, $H_s=\SI{1}{mm}$, $L_I = \SI{3}{mm}$, $L_O = \SI{2}{mm}$,  and set $L$, $H$, $H_F/(H_F+ H_B)$ to be the same as in experiments. Finally, when compared with experiments, we set $\sigma$ based on the charge regulation model for glass \cite{Behrens2001TheSurfaces} immersed in solution with ionic composition $c_k^{in}$. Although we assume uniform surface charge $\sigma$ for the charged channel, in experiments $\sigma$ may vary significantly in the macroporous material due to charge regulation with local concentration and pH. This will be considered in future work. In this paper, $\sigma$ is treated like an artificial parameter for the model. All the constants used in this paper are summarized in Table \ref{tab:constants}, and parameters that vary for different models will be declared where appropriate.

In the following, we show simulation results for a binary electrolyte in Sec.\ref{sec:binary}. The transport of hydronium and hydroxide is also included in addition to the salt ions.

\subsection{Results}{\label{sec:binary}}
For this part, we choose a base case: $\SI{10}{mM}$ NaCl aqueous solution at pH 7 is injected into the feed and electrode channels, and $\sigma = -\SI{20.8}{mC/m^2}$, $h_p=\SI{250}{nm}$, $H = \SI{2.7}{mm}$, $Q = \SI{76}{\mu L/min}$, $Q_E = 4 Q$, $H_F/(H_F+H_B)=0.45$. We first present the results for the base case and identify the importance of hydronium transport and electroosmosis for shock ED. Next, we obtain results for different working conditions and compare the results with available experimental data. Finally, we investigate more parameters that have not yet been tested experimentally, and give suggestions for optimization and scale-up.

In the following, we calculate current efficiency $\eta$, dimensionless overlimiting conductance $\tilde{\kappa}$, dimensionless fresh concentration $\tilde{c}^F$, water recovery $\omega$, and specific energy consumption $\mathcal{E}$ to evaluate the shock ED process. The current efficiency $\eta$ is defined as the current carried by the salt cations divided by the total current through the membranes. The overlimiting conductance $\kappa$ is estimated by $\Delta I/\Delta V$ on the $I$-$V$ curves over the range $\tilde{I} = 2$--$4$. Then it is scaled by the theoretical surface conductance  $\kappa_{sc} = z_+D_+ |\sigma| \epsilon_p h_{io} L/V_T H h_p$ \cite{Dydek2011OverlimitingMicrochannel},
where "+" represents the salt cation, and $\epsilon_p h_{io} L$ is the total cross-sectional area of the charged channels or pores in the $xz$-plane. The fresh ion concentration $c_k^F$ is defined as the flow-weighted mean concentration of the fresh stream. We set $c^F=c_+^F$  to represent the fresh concentration of the binary electrolyte, and scale it by $c_+^{in}$. The water recovery is defined as $\omega = Q_F/Q$, where $Q_F$ is  the fresh stream flow rate. Finally, define the specific energy consumption $\mathcal{E} = (IV + Q \Delta p)/\omega Q$, where $\Delta p$ is the pressure drop across the device and the pump efficiency is assumed to be 1.

\subsubsection{Hydronium transport and electroosmotic vortices}

\begin{figure*}
    \centering
    \includegraphics[width =  1 \textwidth]{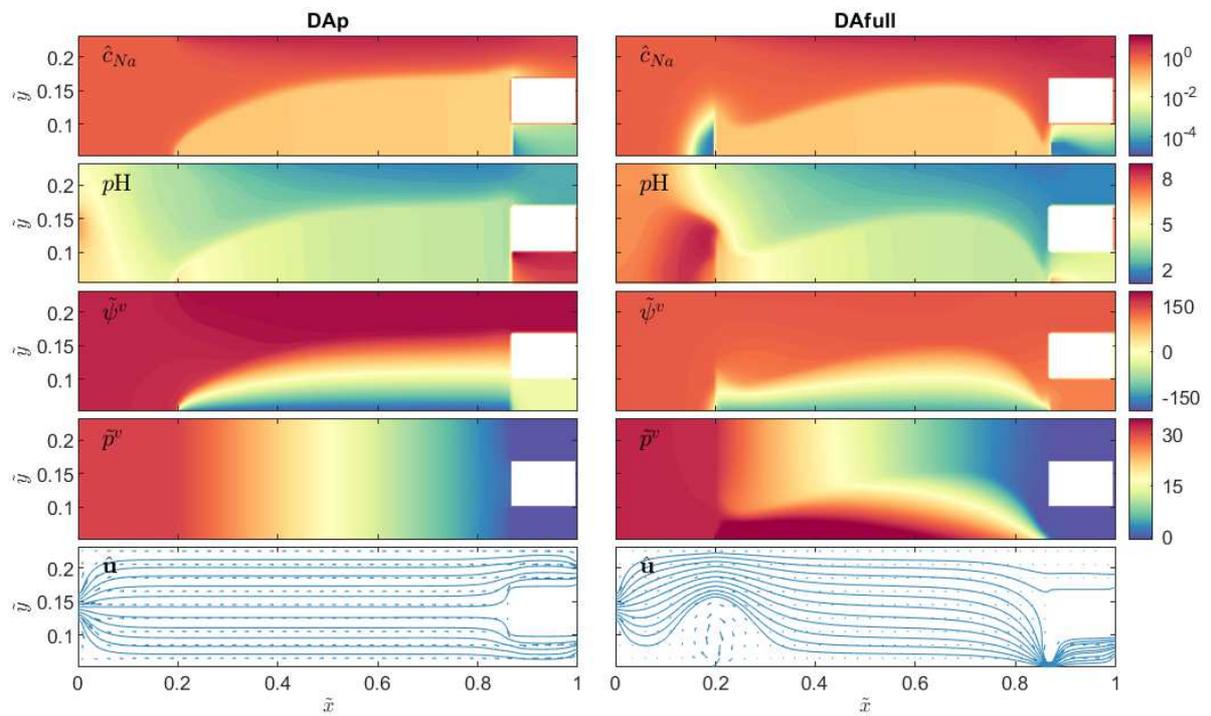}
    \caption{ Steady-state profiles of concentration $\hat{c}_{Na}$, pH, electric potential $\tilde{\psi}^v$, pressure $\tilde{p}^v$, and velocity $\hat{\mathbf{u}}$ at $\tilde{I}=3$ for shock ED. The electrode channels and the membranes are not shown here. The left column corresponds to the DAp model, while the right column corresponds to the DAfull model.  The feed to the main channel and the electrode channels is $\SI{10}{mM}$ NaCl aqueous solution with $p\mathrm{H}=7$. The white blocks represent the splitter.}
    \label{fig:profiles}
\end{figure*}

\begin{figure}[h]
    \centering
    \includegraphics[width = 0.3\textwidth]{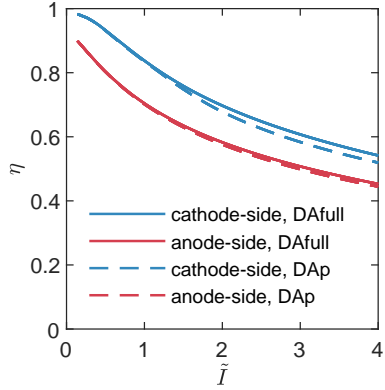}
    \caption{The current efficiency $\eta$ through each of the membranes for the base case shown in Fig.\ref{fig:profiles}.}
    \label{fig:eff_B}
\end{figure}

Fig.\ref{fig:profiles} shows the profiles of $\hat{c}_{Na}$, pH, $\tilde{\psi}^v$, $\tilde{p}^v$ and $\hat{\mathbf{u}}$  for the base case. Only the feed channel is shown. As we can see, for both models, an ion-depleted zone forms near the cathode-side in the charged channel, where almost all of the potential drop occurs. In the depleted zone, the concentration of Na$^+$ is almost constant, while that of H$^+$ increases in the $x$-direction, which should be due to the downward transport of H$^+$ released at the anode. This is the first time that pH profiles of shock ED are shown by modeling. As we can see, the pH in the charged channel is usually below 5, which indicates that the transport of OH$^-$ is negligible compared with H$^+$, and the concentration of H$^+$ is comparable with that of Na$^+$ in the ion-depleted zone. More hydronium ions are produced at the anode as current increases, which results in decrease of current efficiency $\eta$ through the membranes especially at the anode side, as also shown in Fig.\ref{fig:eff_B}.

\begin{figure}
    \centering
    \includegraphics[width = 0.5 \textwidth]{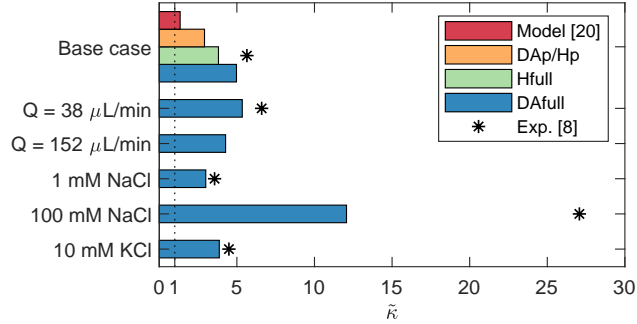}
    \caption{The conductance scaled by the theoretical surface conductance ($\tilde{\kappa} = \kappa / \kappa_{sc}$) at overlimiting current from different models and experimental data \cite{Schlumpberger2015ScalableElectrodialysis}.  }
    \label{fig:cond}
\end{figure}
\begin{figure*}
    \centering
    \includegraphics{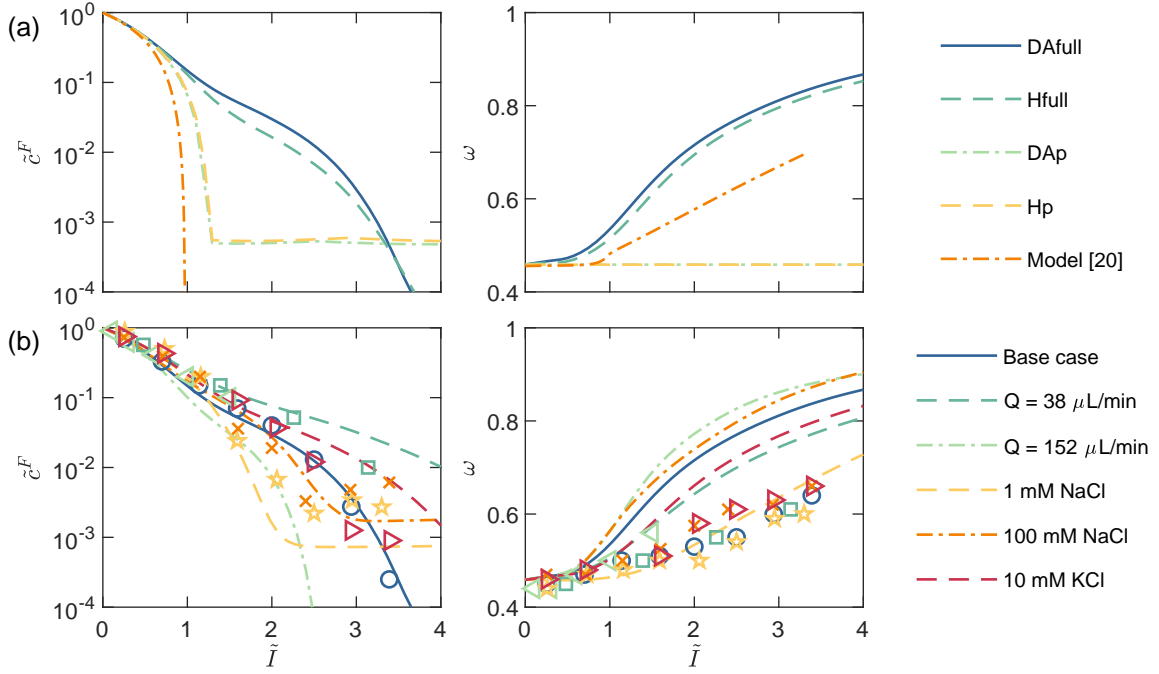}
    \caption{The dimensionless fresh concentration $\hat{c}^F$ (left column) and water recovery $\omega$ (right column) as functions of dimensionless current $\tilde{I}$. (a) Comparison of different models. (b) Comparison of the DAfull model (curves) with experimental data \cite{Schlumpberger2015ScalableElectrodialysis} (markers with the same color as corresponding curves for each case). }
    \label{fig:desal_wr}
\end{figure*}
\begin{figure}
    \centering
    \includegraphics[width = 0.5 \textwidth]{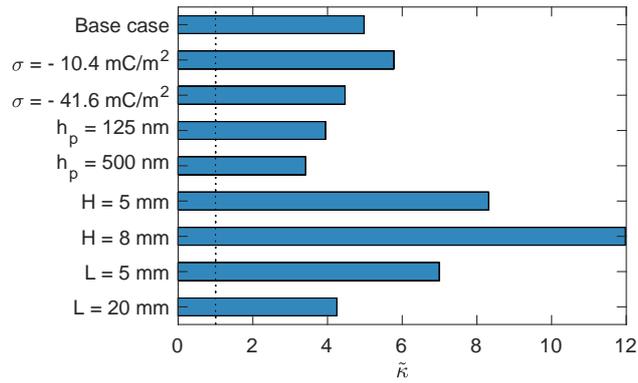}
    \caption{The conductance scaled by the theoretical surface conductance ($\tilde{\kappa} = \kappa / \kappa_{sc}$) at overlimiting current from the DAfull model, for different surface charge density $\sigma$, pore size $h_p$, and macroscopic geometries $H$ and $L$ of the charged channels. }
    \label{fig:conductance_app}
\end{figure}
\begin{figure*}
    \centering
    \includegraphics{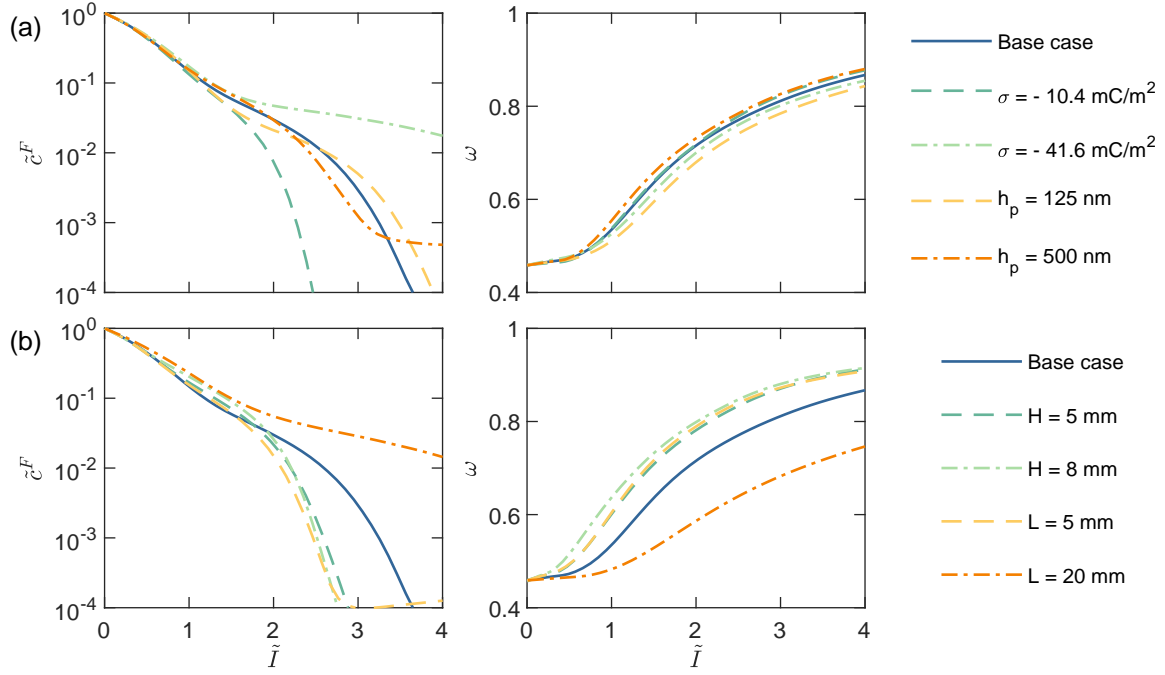}
    \caption{The dimensionless fresh concentration $\hat{c}^F$ (left column) and water recovery $\omega$ ( right column) as functions of dimensionless current $\tilde{I}$ from the DAfull model, for (a) varied  surface charge density $\sigma$ and pore size $h_p$, and (b) varied macroscopic geometries $H$ and $L$ of the charged channels. }
    \label{fig:desal_wr_app}
\end{figure*}
\begin{figure*}
    \centering
    \includegraphics[width = 1 \textwidth]{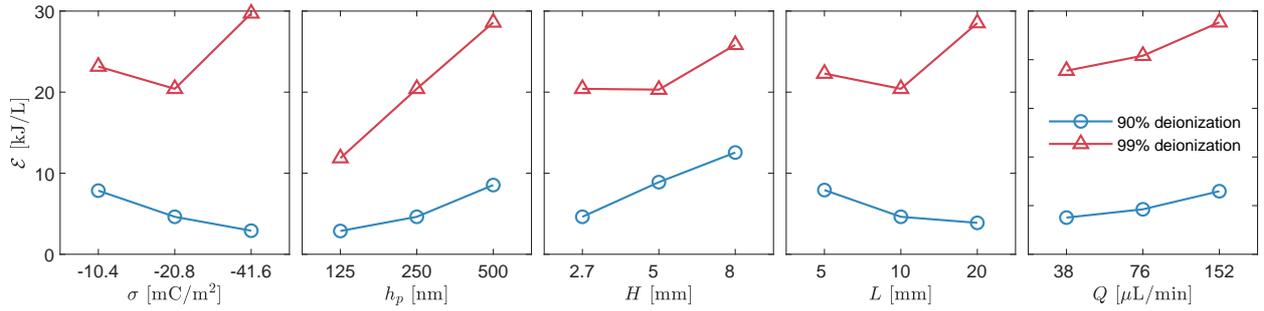}
    \caption{The specific energy consumption $\mathcal{E}$ for 90\% deionization and 99\% deionization of $\SI{10}{mM}$ NaCl solution by shock ED. For each case, only one parameter of $\sigma$, $h_p$, $H$, $L$ and $Q$ is changed relative to the base case. }
    \label{fig:energy}
\end{figure*}
\begin{figure}
    \centering
    \includegraphics[width=0.5\columnwidth]{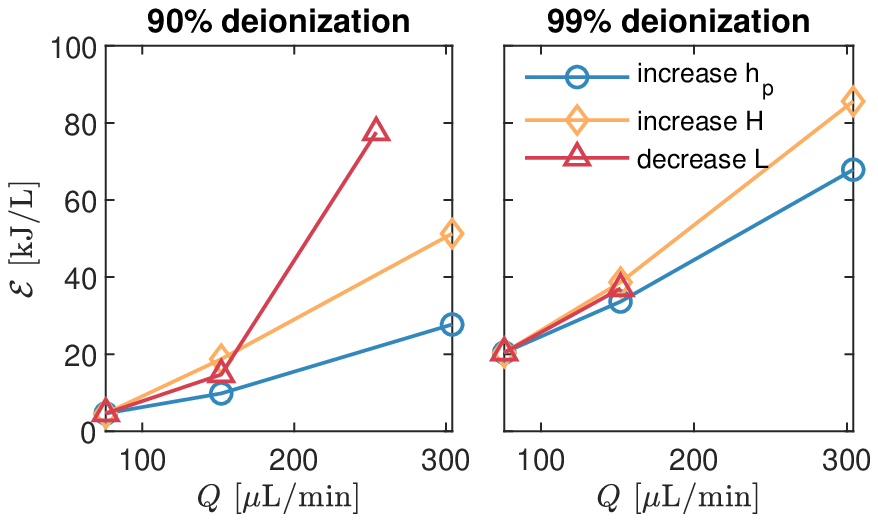}
    \caption{The specific energy consumption $\mathcal{E}$ for 90\% deionization and 99\% deionization of $\SI{10}{mM}$ NaCl solution by shock ED. For each case, the feed flow rate $Q$ and one parameter of $h_p$, $H$, and $L$ are changed relative to the base case, while $Q L /h_p^2 H$ is kept as a constant.}
    \label{fig:scale_up}
\end{figure}
We now examine the roles of electroosmosis and diffusioosmosis on shock ED. First, we calculate $(||u^{DO}|| + ||v^{DO}||)/(||u^{EO}|| + ||v^{EO}||)$ for the velocity field obtained from the DAfull model. This value is less than $0.1\%$, which means that electroosmosis overwhelms diffusioosmosis. In the charged channel, the electroosmotic flow tends to be aligned with the electric field but is hindered by the membrane, which will build up a pressure near the cathode side to push the fluid either up or to the two sides to the inlet and outlet zones. Therefore, two vortices are formed near the interfaces between the charged channel and the inlet and outlet zones. The vortices bring the brine down to the depleted zone and bring the fresh water up in the inlet and outlet zones. Therefore, while the depleted zone grows monotonically in the $x$-direction in the DAp model, the depleted zone shrinks at the outlet in the DAfull model. This can make it extremely hard for the upper boundary of the depleted zone to pass the splitter. As current increase, most of the flow leaves the charged channel through the depleted zone, which guarantees strong desalination, but the ion diffusion from the brine zone to the outlet zone may increase the concentration of the fresh product. We quantitatively analyze deionization later. 

Next, we compare the dimensionless overlimiting conductance $\tilde{\kappa} = \kappa/\kappa_{sc}$ from different models  for the basic case, as shown in Fig.\ref{fig:cond}.  As we can see, the conductance from the model of \cite{Schlumpberger2020shockEDcross} is very close to the theoretical surface conductance. With the adding of hydronium transport (note that $D_{\mathrm{H}^+} \approx 7 D_{\mathrm{Na}^+}$), the DAp model and Hp model both increases $\tilde{\kappa}$ to about 1.5. With the electroosmotic vortices, the Hfull and DAfull model predict even higher conductance. Between these two, the DAfull model gives higher conductance because of the overall more negative zeta potential.

We now consider the fresh concentration $\tilde{c}_k^F$ and water recovery $\omega$ of the fresh stream for the base case, as shown in Fig.\ref{fig:desal_wr}(a). For the models in this paper, $c^F = c_{Na}^F$. For the model of \cite{Schlumpberger2020shockEDcross}, $c^F = c_-^F$ (the authors use the anion concentration in the frit to represent the outlet salt concentration, since they do not have unsupported electrolyte in inlet and outlet zones for simplicity). As we can see, the model of  \cite{Schlumpberger2020shockEDcross} predicts almost perfect desalination at $\tilde{I} \approx 1$. With hydronium transport included, the DAp and Hp model both show less ion removal at the same $\tilde{I}$, and the curve reaches a plateau since the membrane is not ideal and the deionization shock passes the splitter. The fresh concentration decreases most slowly with current for the DAfull and Hfull model, which should be the result of a thin deionization zone near the outlet zone induced by electroosmotic vortices. In addition, the DAfull and Hfull models predict greater water recovery than the model from \cite{Schlumpberger2020shockEDcross}, while DAp and Hp models cannot predict an increase in water recovery with current. The small differences between DAp and Hp as well as DAfull and Hfull indicate that the homogenized model works well for shock ED with a binary electrolyte.

\subsubsection{Comparison with experiments}{\label{subsubsec:expB}}

In this part, we use the DAfull model to calculate additional cases with different working conditions relative to the base case, and we compare the overlimiting conductance, fresh concentration, and water recovery with experimental results \cite{Schlumpberger2015ScalableElectrodialysis}. The new working conditions are as follows: change the feed flow rate $Q$ to 38 or $\SI{152}{\mu L/min}$, change the concentration of NaCl to 1 or $\SI{100}{mM}$ (the corresponding surface charge densities are $\sigma = -10.4$, $-\SI{38.4}{mC/m^2} $), or change NaCl to KCl.  

To begin with, we compare the overlimiting conductance. Note that for the case of $Q = \SI{152}{\mu L/min}$, experimental data \cite{Schlumpberger2015ScalableElectrodialysis} for $\tilde{I}>2$ are not available and thus not shown in Fig.\ref{fig:cond}. As we can see, the experimental overlimiting conductance is much larger than the theoretical surface conductance $\kappa_{sc}$. With the consideration of hydronium transport and electroosmotic vortices, the DAfull model predicts very close results to the experimental data except for the case of $\SI{100}{mM}$, when the magnitude of current is much larger than other cases and the membrane may lose some selectivity in experiments. We can also conclude that electroosmotic vortices or hydronium transport contributes more to the conductance for higher inlet concentrations and smaller flow rates.

As shown in Fig.\ref{fig:desal_wr}(b), the DAfull model provides good predictions for the fresh concentration $\tilde{c}^F$. For $\tilde{I}<1$, almost all the cases converge to a single line. For $\tilde{I}>1$, the model and experiments show that high flow rates favor deionization, and  $\tilde{c}^F$ first decreases rapidly and then reaches a plateau at about $\SI{1e-3}{}$ for the cases of $\SI{1}{mM}$ and $\SI{100}{mM}$. The $\tilde{c}^F$ is smallest at medium concentration at $\tilde{I}=4$, which could be explained by the facts that $\tilde{c}_s$ is large for small feed concentration and the membrane loses selectivity for high concentration, which disfavor deionization. We also expect shock ED to have slightly better deionization of NaCl compared with KCl at high current, which is not apparent in experiments. On the other hand, the DAfull model overestimates the water recovery. One possible reason is that the heterogeneous porous structure of the frit is important for the electroosmotic flow \cite{Mirzadeh2020VorticesMedia}, which is not captured by our planar shock ED model. 



\subsubsection{Optimization and scale-up}
In this part, we first give a primary investigation on how to optimize the shock ED process in terms of specific energy consumption and deionization, based on the current prototype design. Four new parameters are investigated here: the surface charge $\sigma$, the pore size $h_p$, and the length $L$ and the width $H$ of the charged channels (porous material in experiments). We change each parameter twice based on the base case, and get 8 new cases. The scaled overlimiting conductance, fresh concentration, and water recovery for these new cases are shown in Fig.\ref{fig:conductance_app} and \ref{fig:desal_wr_app}. The flow rate cases in Sec.\ref{subsubsec:expB} are also re-visited in this part.  The $\mathcal{E}$  at these two deionization levels ($1-\tilde{c}^F$)  along with the above five parameters is shown in Fig.\ref{fig:energy}. It turns out that the pumping energy represents less than 10\% of the total energy consumed at 90\% deionization, and less than 2\% at 99\% deionization. So the electrical energy dominates energy consumption. In addition, the specific electrical energy consumption scales as $IV/\omega Q \sim \tilde{I}^2 Q/\omega \kappa_{sc}\tilde{\kappa}$ at over-diffusion-limiting current with a constant prefactor for all the cases considered in this part.    As we can see in Fig.\ref{fig:energy}, if the aim is to remove 90\% ions, we should choose more negative $\sigma$, smaller $h_p$, $H$, $Q$, and larger $L$ to reduce the specific energy consumption. The main reason is that 90\% deionization is obtained at around $\tilde{I} = 1$ for all the cases (see Fig.\ref{fig:desal_wr}, \ref{fig:desal_wr_app}), while the above parameter set leads to smaller $\tilde{I}^2 Q/\omega \kappa_{sc} \tilde{\kappa}$ (recall that $\kappa_{sc} \sim \frac{|\sigma|L}{H h_p}$ and see Fig.\ref{fig:conductance_app} for $\tilde{\kappa}$ and Fig.\ref{fig:desal_wr_app} for $\omega$). However, if we aim to remove 99\% ions (obtained at $\tilde{I}>2$), the trends can be different. More negative $\sigma$ leads to less deionization at the same $\tilde{I}$ but also larger overlimiting conductance, so medium $\sigma$ should require the least $\mathcal{E}$.  Larger $H$ leads to more water recovery and deionization at the same $\tilde{I}$ but also smaller conductance, so there is a plateau of $\mathcal{E}$ between $H = 2.7$--$\SI{5}{mm}$. In contrast, larger $L$ leads to larger conductance but less water recovery and deionization, so medium $L$ would be the most favorable. To conclude, to obtain 99\% deionization at the minimal energy cost, we need the macroporous material to have small $h_p$ and moderate $\sigma$, $H$ and $L$, and operate the system under flow feed rate $Q$. 
However, note that the decrease of $h_p$ requires higher pressure at the inlet and thus more robust sealing. 

To scale up the output rate without increasing the pumping pressure, we increase $H$, increase $h_p$, or decrease $L$ to increase $Q$ and keep the pressure drop $\Delta p$ (which $\sim QL/h_p^2 H$) at zero current unchanged. The specific energy consumption at two deionization levels is shown in Fig.\ref{fig:scale_up}. So six new cases are added. Basically, as we scale up the process, the specific energy consumption increases. Among the three parameters, the scale-up by $h_p$ increases the specific energy in the least extent. When we decrease $L$,  the current density becomes very high at 90\% deionization, and transport in electrode streams get limited by diffusion, so the voltage and energy consumption becomes very large. A numerical problem occurs when we want to reach more deionization, so the result for the smallest $L$ at 99\% deionization is not shown in the figure.  In the future, we aim to create more innovative designs, instead of just varying geometries,  for scale-up of the process without sacrificing much of the energy consumption.


%% file: 5_discussion.tex
\section{Conclusion}

This work established a depth-averaged model for shock electrodialysis with multiple ions and more realistic boundary conditions. We identified the importance of hydronium transport and electroosmotic vortices for shock ED performance in terms of conductance AND deionization, and find good consistency between the model and experiments for binary electrolyte. We also give suggestions on how to optimize the process in terms of specific energy consumption, and how to improve the flow rate without increasing pressure, based on the current prototype design. In the second part of the series paper, we will give a preliminary analysis for selective ion removal. In the future, we will also investigate the effect of adsorption and porous heterogeneity on shock ED.

%% file: appendix.tex
\appendix

\section{Coefficients of the depth-averaged model}{\label{sec:appendix}}
In this part, we introduce how we calculate the coefficients $\delta_k$, $\alpha^{EO}$, $\alpha^{DO}_k$, $\beta_k^P$, $\beta_k^{EO}$, and $\beta_{kl}^{DO}$. First, we need to solve the PB equation (Eq.(\ref{eq:PB2})) for $\varphi$ with given $c_k^v$, $\sigma$, and $h$. Then we can use the $\varphi$ to calculate the coefficients: 
 \begin{equation}
    \delta_k =\frac{\overline{c_k}}{c_k^v} = \frac{1}{h} \int_0^h \exp(-z_k \tilde{\varphi}) dz.
\end{equation}
\begin{equation}
    \alpha^{EO} = \frac{1}{h}\int_0^h \left(1 -\frac{\varphi}{\zeta} \right) dz,
\end{equation}
\begin{equation}
    \alpha_k^{DO} = \frac{1}{h}\int_0^h \chi_k(x, y, z) d z,
\end{equation}
where 
\begin{equation}
    \chi_k(x,y,z) = \frac{1}{h^2}\int_z^h \int_0^{z''} \left[\exp(-z_k \tilde{\varphi}') -1\right]dz' dz'',
\end{equation}
and
\begin{equation}
    \beta_k^P = \frac{\overline{\mathbf{u}^Pc_k}}{\overline{\mathbf{u}^P} \cdot \overline{c_k}} = \frac{\frac{1}{h}\int_0^h \left(1 - \frac{z^2}{h^2} \right)\exp(-z_k \tilde{\varphi}) dz}{\frac{2}{3} \delta_k},
\end{equation}
\begin{equation}
    \beta_k^{EO} = \frac{\overline{\mathbf{u}^{EO} c_k}}{\overline{\mathbf{u}^{EO}} \cdot \overline{c_k}}  =  \frac{\frac{1}{h} \int_0^h (1 -\frac{\varphi}{\zeta})  \exp(-z_k \tilde{\varphi}) d z}{\alpha^{EO} \delta_k },
\end{equation}

\begin{equation}
    \beta_{kl}^{DO} = \frac{\overline{\mathbf{u}^{DO}_l c_k}}{\overline{\mathbf{u}^{DO}_l} \cdot \overline{c_k}}  =\frac{\frac{1}{h}\int_0^h \exp(-z_k \tilde{\varphi}) \chi_l dz}{\alpha_l^{DO} \delta_k}.
\end{equation}